\documentclass[twoside]{article}
\usepackage{pgfpages}
\usepackage{jp}
\pgfpagesuselayout{2 on 1 shifted}[border shrink=-1.5in,vertical shift=-.35in,horizontal shift=.175in,in shift=.25in]
\usepackage[utf8]{inputenc}
\usepackage{amssymb}
\usepackage{amsmath}
\usepackage{amsfonts}
\usepackage{color}
\usepackage{graphicx}
\usepackage{bm}
\usepackage{isabelle,isabellesym}
\usepackage{pdfsetup}
\usepackage{float}

\title{Computer Science and Metaphysics:\\ A
  Cross-Fertilization}

\author{Daniel Kirchner\\Fachbereich Mathematik und Informatik \\Freie
  Universit\"at Berlin \\ {\tt daniel@ekpyron.org} \and Christoph
  Benzm\"uller \\ Fachbereich Mathematik und Informatik \\ Freie
  Universit\"at Berlin \\
  {\tt c.benzmueller@fu-berlin.de} \\
  \& 
  Faculty of Science, Technology and Communication \\
  University of Luxembourg  \\ 
  \and Edward
  N. Zalta\\ Center for the Study of Language and
  Information\\ Stanford University \\ {\tt zalta@stanford.edu} }

\date{}

\begin{document}

\maketitle

\begin{abstract}

Computational philosophy is the use of mechanized computational
techniques to unearth philosophical insights that are either difficult
or impossible to find using traditional philosophical
methods. Computational metaphysics is computational philosophy with a
focus on metaphysics.  In this paper, we (a) develop results in modal
metaphysics whose discovery was computer assisted, and (b) conclude
that these results work not only to the obvious benefit of philosophy
but also, less obviously, to the benefit of computer science, since
the new computational techniques that led to these results may be more
broadly applicable within computer science. The paper includes a
description of our background methodology and how it evolved, and a
discussion of our new results.

\end{abstract}

\thispagestyle{empty}

\section{The Basic Computational Approach to \\ Higher-Order Modal Logic}

The application of computational methods to philosophical problems was
initially limited to first-order theorem provers. These are easy to
use and have the virtue that they can do proof discovery.  In
particular, Fitelson and Zalta\footnote{Fitelson \& Zalta, ``Steps
  toward a computational metaphysics''.} both (a) used Prover9 to find
a proof of the theorems about situation and world
theory\footnote{Zalta, ``Twenty-Five basic theorems''.} and (b) found
an error in a theorem about Plato's Forms that was left as an exercise
in a paper by Pelletier \&\ Zalta.\footnote{Pelletier \& Zalta, ``How
  to Say Goodbye to the Third Man''.} And, Oppenheimer and Zalta
discovered,\footnote{Oppenheimer \& Zalta, ``A
  computationally-discovered simplification''.} using Prover9, that 1
of the 3 premises used in their reconstruction of Anselm's ontological
argument\footnote{Oppenheimer \& Zalta, ``On the logic of the
  ontological argument''.} was sufficient to derive the conclusion.
Despite these successes, it became apparent that working within a
first-order theorem-proving system involved a number of technical
compromises that could be solved by using a higher-order system. For
example, in order to represent modal claims, second-order
quantification, and schemata, etc., in Prover9, special techniques
must be adopted that force formulas which are naturally expressed in
higher-order systems into the less-expressive language of multi-sorted
first-order logic. These techniques were discussed in the papers just
mentioned and outlined in some detail in a paper by Alama,
Oppenheimer, and Zalta.\footnote{Alama, Oppenheimer \& Zalta,
  ``Automating Leibniz's theory of concepts''. For example, to
  represent the T schema $\Box \phi \to \phi$, one begins with the
  intermediate representation $\forall p(\Box p \to p)$. Then one
  introduces two sortal predicates {\tt Proposition(x)} and {\tt
    Point(x)}, where the latter represent the possible worlds. In
  addition, a distinguished point {\tt W} must be introduced, as well
  as a truth predicate {\tt True(x,y)}. Then one can represent
  $\forall p(\Box p \to p)$ as: {\tt all x (Proposition(x) -> (all y
    (Point(y) -> True(x,y)) -> True(x,W)))}.} The representations of
expressive, higher-order philosophical claims in first-order logic is
therefore not the most natural; indeed, the complexity of the
first-order representations grows as one considers philosophical
claims that require greater expressivity. And this makes it more
difficult to understand the proofs found by the first-order provers.

\subsection{The Move to Higher-Order Systems}

Ways of addressing such problems were developed in a series of papers
by Benzm\"uller and colleagues. Using higher-order theorem provers,
they brought computational techniques to the study of philosophical
problems and, in the process, they (along with others) developed two
methodologies: \begin{itemize}
  \item Using the syntactic capabilities of a higher-order theorem
    prover such as Isabelle/HOL (a) to represent the semantics of a
    target logic and (b) to define the original syntax of the target
    theory within the prover. We call this technique Shallow Semantic
    Embeddings (SSEs).\footnote{This is to be contrasted with a
      \emph{deep semantic embedding}, in which the syntax of the
      target language is represented using an inductive data structure
      (e.g., following the BNF of the language) and the semantics of a
      formula is evaluated by recursively traversing the data
      structure. \emph{Shallow} semantic embeddings, by contrast,
      \emph{define} the syntactic elements of the target logic while
      reusing as much of the infrastructure of the meta-logic as
      possible.} These SSEs suffice for the implementation of
    interesting modal, higher-order, and non-classical logics and for
    the investigation of the validity of philosophical arguments. By
    proving that the axioms or premises of the target system are true
    in the SSE, one immediately has a proof of soundness of the target
    system.
 \item Developing additional abstraction layers to represent the
   deductive system of philosophical theories with a reasoning system
   that goes beyond the deductive systems of classical modal
   logics.  \end{itemize}
Early papers focused on the development of SSEs. These papers show
that the standard translation from propositional modal logic to
first-order logic can be concisely modelled (i.e., embedded) within
higher-order theorem provers, so that the modal operator $\Box$, for
example, can be explicitly defined by the $\lambda$-term $\lambda
\varphi. \lambda w. \forall v. (R w v \rightarrow \varphi v)$, where
$R$ denotes the accessibility relation associated with $\Box$.  Then
one can construct first-order formulas involving $\Box\varphi$ and use
them to represent and prove theorems. Thus, in an SSE, the target
logic is internally represented using higher-order constructs in an
automated reasoning environment such as Isabelle/HOL.  Benzm\"uller
and Paulson\footnote{Benzm{\"u}ller \& Paulson, ``Quantified
  multimodal logics''.} developed an SSE that captures quantified
extensions of modal logic (and other non-classical logics). For
example, if $\forall x. \varphi x$ is shorthand in functional type theory
for $\Pi (\lambda x. \varphi x)$, then $\Box \forall x Px$ would be
represented as $\Box \Pi' (\lambda x. \lambda w. P x w)$, where $\Pi'$
stands for the $\lambda$-term $\lambda \Phi . \lambda w . \Pi(\lambda
x . \Phi x w)$, and the $\Box$ gets resolved as described
above.\footnote{To see how these expressions can be resolved to
  produce the right representation, consider the following series of
  reductions:\\ \hspace*{.2in}\begin{tabular}{lll} $\Box\forall x P x$
    & $\equiv$ & $\Box \Pi' (\lambda x. \lambda w. P x w)$\\ &
    $\equiv$ & $\Box ((\lambda \Phi . \lambda w . \Pi(\lambda x . \Phi
    x w)) (\lambda x. \lambda w. P x w))$\\ & $\equiv$ & $\Box
    (\lambda w . \Pi(\lambda x . (\lambda x. \lambda w. P x w) x
    w))$\\ & $\equiv$ & $\Box (\lambda w . \Pi(\lambda x . P x
    w))$\\ & $\equiv$ & $(\lambda \varphi. \lambda w. \forall v. (R w
    v \rightarrow \varphi v)) (\lambda w . \Pi(\lambda x . P x
    w))$\\ & $\equiv$ & $(\lambda \varphi. \lambda w. \Pi (\lambda v
    . R w v \rightarrow \varphi v)) (\lambda w . \Pi(\lambda x . P x
    w))$\\ & $\equiv$ & $(\lambda w. \Pi (\lambda v . R w v
    \rightarrow (\lambda w . \Pi(\lambda x . P x w)) v)) $\\ &
    $\equiv$ & $(\lambda w. \Pi (\lambda v . R w v \rightarrow
    \Pi(\lambda x . P x v)) ) $\\ & $\equiv$ & $(\lambda w. \forall v
    . R w v \rightarrow \forall x . P x v) $\\ & $\equiv$ & $(\lambda
    w. \forall v x . R w v \rightarrow P x v) $
  \end{tabular}\\
Thus, we end up with a representation of  $\Box\forall x P x$ in
functional type theory.}

The SSE technique was also the starting point for a natural encoding
of G\"odel's modern variant of the ontological argument in
second-order S5 modal logic.  Various computer formalizations and
assessments of recent variants of the ontological argument in
higher-order theorem provers emerged in work by Benzm\"uller and
colleagues. Initial studies\footnote{Benzm{\"u}ller \& Woltzenlogel-Paleo,
  ``Automating {G\"{o}del's} ontological proof''.} investigated
G\"odel's and Scott's variants of the argument within the higher-order
automated theorem prover (henceforth ATP)
LEO-II.\footnote{Benzm{\"u}ller, Sultana, Paulson \& Thei{\ss}, ``The
  higher-order prover {LEO-II}''.} Subsequent work deepened these
assessment studies.\footnote{Benzm{\"u}ller \& Woltzenlogel-Paleo, ``The
  inconsistency in {G{\"o}del's} ontological
  argument''. Benzm{\"u}ller \& Woltzenlogel-Paleo, ``Object-logic explanation for
  the inconsistency in {G\"odel's} ontological theory''.} Instead of
using LEO-II, these studies utilized the higher-order proof assistant
Isabelle/HOL,\footnote{Nipkow, Paulson \& Wenzel, ``Isabelle/HOL``.} a
system that is interactive and supports strong proof automation. Some
of these experiments were reconstructed in the proof assistant
Coq.\footnote{Bertot \& Casteran, ``Interactive Theorem Proving and
  Program Development``. Benzm{\"{u}}ller \& Woltzenlogel-Paleo, ``Interacting with
  modal logics in the {Coq} proof assistant''.}  Additional follow-up
work contributed similar studies\footnote{Benzm{\"u}ller, Weber \&
  Woltzenlogel-Paleo, ``Analysis of the {Anderson-H\'{a}jek}
  controversy''. Fuenmayor \& Benzm{\"u}ller, ``Automating emendations
  of the ontological argument''. Fuenmayor \& Benzm\"uller, ``A case
  study on computational hermeneutics''. Bentert, Benzm{\"u}ller,
  Streit \& Woltzenlogel-Paleo, ``Analysis of an ontological proof proposed by
  {Leibniz}''.} and includes a range of variants of the ontological
argument proposed by other authors, such as Anderson, H\'ajek, Fitting,
and Lowe.\footnote{Anderson, ``Some emendations of {G{\"o}del's}
  ontological proof''. Anderson \& Gettings, ``{G\"odel}'s ontological
  proof revisited''. H\'ajek, ``Magari and others on {G\"odel’s}
  ontological proof''. H\'ajek, ``{Der Mathematiker und die Frage der
    Existenz Gottes}''. H{\'{a}}jek, ``A new small emendation of
  {G{\"{o}}del's} ontological proof''. Fitting, ``Types, Tableaus, and
  {G}{\"o}del's God''. Lowe, ``A modal version of the ontological
  argument''.} Moreover ultrafilters have been
used\footnote{Benzm{\"u}ller \& Fuenmayor, ``Can computers help to
  sharpen our understanding of ontological arguments?''.} to study the
distinction between extensional and intensional positive properties in
the variants of Scott, Anderson and Fitting. This ongoing work will be
sketched in Section \ref{section:Goedel}.

The other main technique (i.e., the one in the second bullet point
above), was developed by Kirchner and Benzm\"uller to re-implement, in
a higher-order system, the work by Fitelson and
Zalta.\footnote{Fitelson \& Zalta, ``Steps toward a computational
  metaphysics''.}  In order to develop a more general implementation
of Abstract Object Theory (henceforth AOT or `object theory'), it
doesn't suffice to just develop an SSE for AOT.  The SSE of G\"odel's
ontological argument relies heavily on the completeness of
second-order modal logic with respect to Kripke models.  Given these
completeness results, the computational analysis at the SSE level
accurately reflects what follows from the premises of the argument.
Since such completeness results aren't available for AOT with respect
to its Aczel models, some other way of investigating the proof system
computationally is needed. To address this, Kirchner extended the SSE
by introducing the new concept of \emph{abstraction
  layers}.\footnote{Kirchner, ``{Representation and Partial Automation
    of the PLM in Isabelle/HOL}''.} By introducing an additional proof
system (as a higher abstraction layer, on top of the semantic
embedding) that involves just the axioms of the target logic, one can
do automated reasoning in the target logic without generating
artifactual theorems (i.e., theorems of the model that aren't theorems
of the target logic), and without requiring the embedding to be
complete or even provably complete. So the additional abstraction
layer makes interactive and automated reasoning in Isabelle/HOL
possible in a way that is independent of the model structure used for
the semantic embedding itself.  Whereas the SSE serves as a sound
basis for implementing the abstract reasoning layer, the embedding
with abstraction layers provides the infrastructure for a deeper
analysis of the semantic properties of the target logic, such as
completeness. We'll expand upon this theme on several occasions below.

Kirchner reconstructs not only AOT's fundamental theorems about
possible worlds, but arrives at meta-theorems about the correspondence
between AOT's syntactic possible worlds and the semantic possible
worlds used in its models.\footnote{Kirchner, ``{Representation and
    Partial Automation of the PLM in Isabelle/HOL}''.} In particular,
Kirchner shows, using the embedding of AOT in Isabelle/HOL, that for
each syntactic possible world $\bm{w}$ of AOT, there exists a semantic
possible world $w$ in the embedding, such that all and only
propositions derivably true in $\bm{w}$ (using AOT's definition of
truth in possible worlds) are true in $w$, and vice-versa. The shallow
semantic embedding with abstraction layers made it possible to reason
both \emph{within} the target logic itself (i.e., in the higher-level
abstract reasoning layer) and \emph{about} the target logic (i.e.,
using the outer logic of HOL as metalogic and the embedding as a
definition of the semantics of the target logic).

To illustrate, as simply as possible, some of the technical details
involved in the two basic techniques described above, we now turn to
the development of an SSE for propositional modal logic, and show how
an abstraction layer can be added on top of that.

\subsection{Propositional S5 with Abstraction Layers.}\label{example_S5}
Our computational method can be illustrated with the simple example of
a shallow semantic embedding of a propositional S5 logic with
abstraction layers. In order to map modal logic to non-modal
higher-order logic we use Kripke semantics. To that end we introduce a
(non-empty) domain \isa{i} for possible worlds in terms of a type
declaration. In Isabelle/HOL:

\begin{quote}
\isakeyword{typedecl} \isa{i}
\end{quote}

Then we define a type for propositions, which are represented by
functions mapping possible worlds to booleans. The right-hand side of
the following in Isabelle/HOL represents the complete set of
\emph{all} (\isa{UNIV}) functions from type \isa{i} to type
\isa{bool}.  This set is used to define a new \emph{abstract} type
\isa{$\mathrm{o}$}, whose objects are represented by elements of this
set.\footnote{To introduce a new type the representation set has to be
  non-empty. The fact that it is non-empty here can be trivially
  proven, which is indicated by the two dots at the end of the line.}
  
\begin{quote}
\isakeyword{typedef} \isa{$\mathrm{o}$ =     "UNIV::(i$\Rightarrow$bool) set" ..}
\end{quote}
  
Given these definitions we then lift the already defined connectives
of our meta-logic HOL to the newly introduced type \isa{$\mathrm{o}$} of
propositions in the target logic:

\begin{quote}
\isakeyword{lift\_definition} \isa{S5\_not ::
  "$\mathrm{o}\Rightarrow\mathrm{o}$" ("$\bm{\neg}$\_" [54]
  70)}\\ \-\hspace{2em}\isakeyword{is} \isa{"$\lambda p.\; \lambda
  w.\; \neg(p\; w)$" .}
\end{quote}

\begin{quote}
\isakeyword{lift\_definition} \isa{S5\_impl ::
  "$\mathrm{o}\Rightarrow\mathrm{o}\Rightarrow\mathrm{o}$"
  (}\isakeyword{infixl} \isa{"$\bm{\rightarrow}$"
  51)}\\ \-\hspace{2em}\isakeyword{is} \isa{"$\lambda p.\; \lambda
  q.\; \lambda w.\; (p\; w)\; \longrightarrow\; (q\; w)$" .}
\end{quote}

The first defines the new operator \isa{S5\_not} (with the convenient
syntax of a bold negation \isa{$\bm{\neg}$}) on the abstract type
\isa{$\mathrm{o}$} using the given \isa{$\lambda$}-function on the
representation type (functions from possible worlds to booleans).  In
\isa{$\lambda p.\; \lambda w.\; \neg(p\; w)$}, the
\isa{$\lambda$}-bound \isa{$p$} is a function from possible worlds to
booleans (type \isa{i$\Rightarrow$bool}) and \isa{$w$} is a possible
world.  The \isa{$\lambda$}-term maps \isa{$p$} and \isa{$w$} to the
negation of: \isa{$w$} applied to \isa{$p$}.  So it defines a function
of type \isa{(i$\Rightarrow$bool)$\Rightarrow$i$\Rightarrow$bool},
which is exactly the representation type of the desired signature
\isa{$\mathrm{o}\Rightarrow\mathrm{o}$}.  The implication connective
\isa{S5\_impl} with syntax \isa{$\bm{\rightarrow}$} is defined in a
similar manner.

Using the same mechanism the unique operators of modal logic can be
defined in accordance with their Kripke semantics:

\begin{quote}
\isakeyword{lift\_definition} \isa{S5\_box ::
  "$\mathrm{o}\Rightarrow\mathrm{o}$" ("$\bm{\Box}$\_" [62]
  63)}\\ \-\hspace{2em}\isakeyword{is} \isa{"$\lambda p.\; \lambda
  w.\; \forall v.\; p\; v$" .}
\end{quote}

\begin{quote}
\isakeyword{lift\_definition} \isa{S5\_dia ::
  "$\mathrm{o}\Rightarrow\mathrm{o}$" ("$\bm{\Diamond}$\_" [62]
  63)}\\ \-\hspace{2em}\isakeyword{is} \isa{"$\lambda p.\; \lambda
  w.\; \exists v.\; p\; v$" .}
\end{quote}

To formulate statements about our newly defined target logic, we still
have to define what it means for a formula of the target logic to be
\emph{valid}.  There are two options for defining validity, i.e.,
either as truth relative to a designated actual world or as truth in
all possible worlds. Thus, to define validity, we first need a
definition of truth relative to a possible world:

\begin{quote}
\hspace*{-.025in}\isakeyword{lift\_definition} \isa{S5\_true\_in\_world ::
  "$i\Rightarrow\mathrm{o}\Rightarrow$bool" ("[\_ $\models$
    \_]")}\\ \-\hspace{2em}\isakeyword{is} \isa{"$\lambda p.\; \lambda
  w.\; p\; w$" .}
\end{quote}

It turns out that this is sufficient for reasoning about validity; so
we don't need to choose between the following two alternative
definitions\footnote{It has been argued that the second option is more
  philosophically correct, see Zalta, ``Logical and analytic truths
  that are not necessary''.} of global validity:

\begin{quote}
\isakeyword{lift\_definition} \isa{S5\_valid\_nec ::
  "$\mathrm{o}\Rightarrow$bool"
  ("$\Box$[\_]")}\\ \-\hspace{2em}\isakeyword{is} \isa{"$\lambda p.\;
  \forall w.\; p\; w$" .}
\end{quote}

\begin{quote}
\isakeyword{consts} \isa{$w_0$ :: i}
\end{quote}

\begin{quote}
\isakeyword{lift\_definition} \isa{S5\_valid\_act ::
  "$\mathrm{o}\Rightarrow$bool"
  ("$\mathcal{A}$[\_]")}\\ \-\hspace{2em}\isakeyword{is}
\isa{"$\lambda p.\; p\; w_0$" .}
\end{quote}

What we have so far is a shallow semantical embedding of an S5 modal
logic, implemented using an \emph{abstract type} in
Isabelle/HOL.\footnote{For the shallow semantical embedding alone we
  could have skipped the introduction of a new abstract type
  \isa{$\mathrm{o}$}, but instead used the representation type
  \isa{i$\Rightarrow$bool} directly in the definitions; however, using
  the abstract type makes it easier to introduce the \emph{abstraction
    layer} in the following paragraphs.}  We can already formulate and
prove statements in our target logic at this stage by initiating what
Isabelle/HOL calls a \emph{transfer} of a given statement to its
counterpart with respect to the representation types, in accordance
with the lifting definitions.\footnote{Huffman \& Kuncar, ``Lifting
  and transfer''.} So to prove the K$\Diamond$-lemma, we simply give
the following command:

\begin{quote}
\isakeyword{lemma} \isa{"[$w \models \bm{\Box}(p \bm{\rightarrow} q)
    \bm{\rightarrow} (\bm{\Diamond}p \bm{\rightarrow} \bm{\Diamond}
    q)$]"}\\ \-\hspace{2em}\isakeyword{apply} \isa{transfer}
\isakeyword{by} \isa{auto}
\end{quote}

The proof of this lemma uses the \isa{transfer} method and is shown to
be valid in the semantics, so the proof doesn't reveal which
particular axioms, or axiom system, of S5 are needed to derive it in
the traditional sense.  In the case of propositional S5 modal logic
this doesn't constitute a problem, since it is known that it is
\emph{complete} with respect to Kripke semantics. So everything that
is derivable from the semantics will also be derivable from the
standard axioms of S5. However, for more complex target systems like
AOT, this is not the case \emph{a priori}.

We could, at this point, show how the additional abstraction layers
needed for the proof theory of AOT can be developed, but that would
introduce complexity that isn't really needed for this discussion. So,
instead, we shall illustrate how an abstraction layer can be added to
the above SSE for propositional modal logic. For the remainder of this
section, then, we proceed as if the completeness results for Kripke
semantics aren't known.  For the analysis of the proof theory of
propositional S5 logic, the first step is to simultaneously show that
the system of propositional S5 logic is \emph{sound} with respect to
our semantics and construct the basis of our abstraction layer by
deriving the standard S5 axioms from the semantics:

\begin{quote}
\isakeyword{lemma} \isa{ax\_K: "[$w \models \bm{\Box}(p
    \bm{\rightarrow} q) \bm{\rightarrow} (\bm{\Box}p \bm{\rightarrow}
    \bm{\Box} q)$]"}\\ \-\hspace{2em}\isakeyword{apply} \isa{transfer}
\isakeyword{by} \isa{auto}\\ \isakeyword{lemma} \isa{ax\_T: "[$w
    \models \bm{\Box}p \bm{\rightarrow}
    p$]"}\\ \-\hspace{2em}\isakeyword{apply} \isa{transfer}
\isakeyword{by} \isa{auto}\\ \isakeyword{lemma} \isa{ax\_5: "[$w
    \models \bm{\Diamond}p \bm{\rightarrow}
    \bm{\Box}\bm{\Diamond}p$]"}\\ \-\hspace{2em}\isakeyword{apply}
\isa{transfer} \isakeyword{by} \isa{auto}

\end{quote}

Furthermore we need axioms for the classical negation and implication
operators, e.g.,

\begin{quote}
\isakeyword{lemma} \isa{ax\_pl\_1: "[$w \models p \bm{\rightarrow} (q
    \bm{\rightarrow} p)$]"}\\ \-\hspace{2em}\isakeyword{apply}
\isa{transfer} \isakeyword{by} \isa{auto}\\
\end{quote}

\noindent and so on for the other axioms of propositional logic.

Next we need to derive the two inference rules, i.e., modus ponens and
necessitation.

\begin{quote}
\isakeyword{lemma} \isa{mp:} \isakeyword{assumes} \isa{"[$w \models
    p$]"} \isakeyword{and} \isa{"[$w \models p \bm{\rightarrow}
    q$]"}\\ \-\hspace{2em}\isakeyword{shows} \isa{"[$w \models
    q$]"}\\ \-\hspace{2em}\isakeyword{using} \isa{assms}
\isakeyword{apply} \isa{transfer} \isakeyword{by} \isa{auto}
\end{quote}

\begin{quote}
\isakeyword{lemma} \isa{necessitation:} \isakeyword{assumes}
\isa{"$\forall v.\quad$[$v \models
    p$]"}\\ \-\hspace{2em}\isakeyword{shows} \isa{"[$w \models
    \bm{\Box}p$]"}\\ \-\hspace{2em}\isakeyword{using} \isa{assms}
\isakeyword{apply} \isa{transfer} \isakeyword{by} \isa{auto}
\end{quote}

Unfortunately, in our implementation we are lacking \emph{structural
  induction}, i.e., induction on the complexity of a formula. For that
reason, we also have to derive meta-rules for our target system from
the semantics, e.g.,

\begin{quote}
\isakeyword{lemma} \isa{deduction:} \isakeyword{assumes} \isa{"[$w
    \models p$] $\Longrightarrow$ [$w \models
    q$]"}\\ \-\hspace{2em}\isakeyword{shows} \isa{"[$w \models p
    \bm{\rightarrow} q$]"}\\ \-\hspace{2em}\isakeyword{using}
\isa{assms} \isakeyword{apply} \isa{transfer} \isakeyword{by}
\isa{auto}
\end{quote}

Together, the axioms, the inference rules and our meta-rules now
constitute the abstraction layer in our embedding. Subsequent
reasoning can be restricted so that it doesn't use the semantic
properties of the embedding (i.e., so that it won't transfer abstract
types to representation types or unfold the semantic definitions). In
this way, proofs can be constructed that only rely on the axioms and
rules themselves.

Given a sane choice of inference rules and meta-rules, every theorem
derived in this manner is guaranteed to be derivable from the axiom
system. While simple propositional S5 modal logic is known to be
complete with respect to its semantic representation, one can still
construct abstraction layers to reproduce and analyze the deductive
reasoning system of a particular formalization of S5. The abstraction
layer can help a user in interactive reasoning, since it enables the
same mode of reasoning as the target system with identical rules. More
generally, whenever the focus of an investigation is derivability
rather than semantic truth,\footnote{This is generally the case when
  investigating an entire logical theory, rather than a logical
  argument that might not even specify a fixed axiom system against
  which it is formulated.} introducing abstraction layers is either
necessary (if there are no completeness results) or at least helpful
(even if there are completeness results), since they alleviate the
need for a translation process from semantic facts to actual
derivations.

A reasonable analysis of AOT is not possible without abstraction
layers.  For one, AOT is more expressive than propositional modal
logic and uses foundations that are fundamentally different from
HOL.\footnote{AOT is formulated in relational type theory, whereas HOL
  is based on functional type theory. A translation between the two is
  known to be challenging, e.g. see Oppenheimer \& Zalta, ``Relations
  versus functions''.} Therefore a representation of the semantics and
a model structure of AOT in HOL is more complex and it becomes more
difficult to reason about AOT solely by unfolding semantic
definitions. Furthermore, there are as yet no results about the
completeness of the canonical \emph{Aczel models} of AOT, so there is
no guarantee that theorems valid in the semantic embedding are in fact
derivable using the axioms and derivation system of AOT itself.
Lastly, although the original motivation for constructing an SSE of
AOT was mainly to investigate the feasibility of a translation between
functional and relational type theory and to gain insights about
possible models of the theory, it turned out that an SSE with
abstraction layers can be used as a means to analyze the effects of
variations in the axiomatization of AOT itself. This led to an
evolution of AOT, parts of which are described in
Section~\ref{AOTResults}.

The remainder of the paper is structured as follows. In Section 2, we
explain how the SSE technique led to insights about G\"odel's
ontological argument.  In Section 3, we discuss the various insights
into AOT that emerged as a result of the addition of the abstraction
layer to the SSE for AOT.  Finally, in Section 4, we discuss how our
techniques may be generalized, and how cross-pollination between
computer science and philosophy works to the benefit of both
disciplines.

\section{Implementation of G\"odel's Ontological\\ Argument}  \label{section:Goedel}

This section outlines the results of a series of experiments in which
the SSE approach was successfully utilized for the computer-supported
assessment of modern variants of the ontological argument for the
existence of God. The first series of experiments, conducted by
Benzm\"uller and Woltzenlogel-Paleo, focused on G\"odel's higher-order
modal logic variant,\footnote{G\"odel, ``Appendix A. Notes in Kurt
  G\"odel's Hand''.} as emended by Dana Scott\footnote{Scott,
  ``Appendix B: Notes in Dana Scott's Hand''.} and others; the
detailed results were presented in the
literature.\footnote{Benzm{\"u}ller \& Woltzenlogel-Paleo, ``The inconsistency in
  {G{\"o}del's} ontological argument''. Benzm{\"u}ller \& Woltzenlogel-Paleo,
  ``Automating {G\"{o}del's} ontological proof''. Benzm{\"u}ller,
  Weber \& Woltzenlogel-Paleo, ``Analysis of the {Anderson-H\'{a}jek}
  controversy''. Fuenmayor \& Benzm\"uller, ``Types, Tableaus and
  G{\"o}del's God''.} This work had a strong influence on the research
mentioned above, since its success motivated the question of whether
the SSE approach would eventually scale for more ambitious and larger
projects in computational metaphysics. The computer-supported
assessments of G\"odel's version of the ontological argument and its
variants revealed several novel findings some of which will be
outlined below.

\subsection{Inconsistency and Other Results about  G\"odel's Argument}
In the course of experiments,\footnote{Benzm{\"u}ller \& Woltzenlogel-Paleo,
  ``Automating {G\"{o}del's} ontological proof''.} the theorem prover
Leo-II detected that the \emph{unedited} version of G\"odel's
formulation of the argument\footnote{G\"odel, ``Appendix A. Notes in
  Kurt G\"odel's Hand''.} was inconsistent, and that the emendation
introduced by Scott\footnote{Scott, ``Appendix B: Notes in Dana
  Scott's Hand''.} while transcribing the original notes was essential
to preserving consistency. The Scott version was verified for logical
soundness in the interactive proof assistants
Isabelle/HOL\footnote{Nipkow, Paulson \& Wenzel, ``Isabelle/HOL``.}
and Coq.\footnote{Bertot \& Casteran, ``Interactive Theorem
  Proving''.}  In Figures 2 and 3, the axioms causing the
inconsistency in G\"odel's manuscript are highlighted. The
inconsistency, which was missed by philosophers, is explained in
detail in related publications.\footnote{Benzm{\"u}ller \& Woltzenlogel-Paleo,
  ``Object-logic explanation for the inconsistency in {G\"odel's}
  ontological theory''. Benzm{\"u}ller \& Woltzenlogel-Paleo, ``The inconsistency
  in {G{\"o}del's} ontological argument''.}

The problem G\"odel introduced in his
\emph{scriptum}\footnote{G\"odel, ``Appendix A. Notes in Kurt
  G\"odel's Hand''.}  is that \emph{essence} is defined
as:\begin{itemize}
  \item A property $Y$ is the \emph{essence} of an individual $x$ iff
    all of $x$’s properties are entailed by $Y$, i.e., iff $\forall
    Z(Zx \to Y\! \Rightarrow\! Z)$, \end{itemize}
where $Y\! \Rightarrow \!Z$ means that $\Box \forall x(Yx \to Zx)$.
We'll see below that this definition doesn't require that an
individual $x$ exemplify its essence, something we would intuitively
expect of the notion of an essence.  Scott, in contrast, added a
conjunct to the definition of essence:
 \begin{itemize}
  \item A property $Y$ is the \emph{essence} of an individual $x$ iff
    $x$ has property $Y$ and all of $x$’s properties are entailed by
    $Y$.
\end{itemize}
This simple emendation by Scott preserved consistency of the axioms
G\"odel introduced as premises of the argument.

The inconsistency in G\"odel's original version already appears when
the argument is formulated in the quantified modal logic K (with and
without the Barcan formulas), and thus also appears in the stronger
logics KB and S5, which are both extensions of K.\footnote{Though in
  S5, the Barcan formulas are derivable.} By proving a simple lemma,
one can demonstrate how the inconsistency arises. The simple lemma is:
 \begin{description}
  \item[EmptyEssenceLemma] \sloppy An empty property (e.g., being
    non-self-identical) is an essence of any individual $x$.
\end{description}
This lemma, in combination with the other highlighted axioms and
definitions in Figures 2 and 3, implies a
contradiction.\footnote{Benzm{\"u}ller \& Woltzenlogel-Paleo, ``Object-logic
  explanation for the inconsistency in {G\"odel's} ontological
  theory''.} The inconsistency was detected automatically by the ATP
Leo-II,\footnote{Benzm{\"u}ller, Sultana, Paulson \& Thei{\ss}, ``The
  higher-order prover {LEO-II}''.} and in the course of the proof, it
used the fact that an empty property obeys the above lemma to derive
the contradiction.\footnote{It is interesting to note here that during
  the course of its discovery of the inconsistency, Leo-II engaged in
  \emph{blind-guessing}. That is, it used a primitive substitution
  rule to instantiate a predicate quantifier $\forall Y$ with the
  $\lambda$-expression $[\lambda x\: x\neq x]$. This is a method that
  is \emph{not} unification-based. See Andrews, ``On connections and
  higher-order logic''.}

The investigation using ATPs also yielded other noteworthy
results:\begin{itemize}
    \item it determined which axioms were otiose,
    \item it determined which properties of the modal
    operator were required for the argument, and
    \item it determined that the argument, even as emended by Scott,
      implies modal collapse, i.e., that $\varphi \equiv \Box\varphi$,
      so that there can only be models of the premises to the argument
      in which there is exactly one possible world. \end{itemize}
The modal collapse was already noted by Sobel\footnote{Sobel,
  ``G\"odel's ontological proof''. Sobel, ``Logic and Theism''.} but
quickly confirmed by the ATP. One might conclude, therefore that the
premises of G\"odel's argument imply that everything is determined (we
may even say: that there is no free will).

Further variants of G\"odel's argument, in which his premises were
weakened to address the above issues, were proposed by Anderson,
H\'ajek, Fitting, and Bj{\o}rdal.\footnote{Anderson, ``Some
  emendations of {G{\"o}del's} ontological proof''. Anderson \&
  Gettings, ``{G\"odel}'s ontological proof revisited''. H\'ajek,
  ``Magari and others on {G\"odel’s} ontological proof''. H\'ajek,
  ``{Der Mathematiker und die Frage der Existenz
    Gottes}''. H{\'{a}}jek, ``A new small emendation of
  {G{\"{o}}del's} ontological proof''. Fitting, ``Types, Tableaus, and
  {G}{\"o}del's God''. Bj{\o}rdal, ``Understanding {G\"{o}del’s}
  ontological argument''.}  The modal collapse problem was the key
motivation for the contributions of Anderson, H\'ajek, and Bj{\o}rdal
(and many others), and these have also been investigated
computationally.\footnote{Benzm{\"u}ller \& Woltzenlogel-Paleo, ``The modal
  collapse''.}  Moreover, ATPs have even
contributed\footnote{Benzm{\"u}ller, Weber \& Woltzenlogel-Paleo, ``Analysis of the
  {Anderson-H\'{a}jek} controversy''.} to the \emph{clarification of
  an unsettled philosophical dispute} between Anderson and H\'ajek.
In the course of this work, \emph{different notions of quantification}
(\emph{actualist} and \emph{possibilist}) have been utilized and
combined within the SSE approach.\footnote{Benzm{\"u}ller \& Woltzenlogel-Paleo,
  ``Higher-order modal logics''.}

\subsection{Emendations by Anderson and Fitting} 

The emendations proposed by C.~Anthony Anderson\footnote{Anderson,
  ``Some emendations of {G{\"o}del's} ontological proof''. Anderson \&
  Gettings, ``{G\"odel}'s ontological proof revisited''.} and Melvin
Fitting\footnote{Fitting, ``Types, Tableaus, and {G}{\"o}del's God''.}
to avoid the modal collapse are rather distinctive and merit special
consideration. In order to rationally reconstruct Fitting's argument,
an SSE of the richer logic underlying his argument was
constructed. This same SSE was used to reconstruct Anderson's
argument. By introducing the mathematical notion of an ultrafilter,
the two versions of the argument can be compared. This enhanced SSE
technique shows that their variations of the argument are closely
related.

\subsubsection{Anderson's Variant.}
Anderson's central change was to modify a premise that governs the
primitive notion of a \emph{positive} property, which was originally
governed by the axiom: $Y$ is positive if and only if the negation of
$Y$ is non-positive (cf.~axiom A2 in Figure 2 where an \emph{exclusive
  or} is utilized). Anderson suggests that one direction of the
biconditional should be preserved, namely, that:\begin{itemize}
  \item[] If a property is positive, then its negation is not positive.
\end{itemize}
As expected, this has an effect on the argument’s validity, and in
order to render the argument logically valid again, Anderson proposes
modifications to premises governing other notions of the argument ---
in particular, to those governing the definition of essence (which
Anderson revises to \emph{essence}$^*$) and a modified notion of
Godlikeness (\emph{Godlike}$^*$):\begin{description}
  \item[essence$^*$] A property $E$ is an essence$^*$ of an individual
    $x$ if and only if all of $x$’s necessary/essential properties are
    entailed by $E$ and (conversely) all properties entailed by $E$
    are necessary/essential properties of $x$.
  \item[Godlike$^*$] An individual $x$ is $\textit{Godlike}^*$ if and
    only if all and only the necessary/essential properties of $x$ are
    positive, i.e., $G^*x \equiv \forall Y (\Box Yx \equiv P(Y))$.
\end{description}
These two amended definitions render the argument logically valid
again. This was verified computationally.\footnote{Fuenmayor \&
  Benzm\"uller, ``Types, Tableaus and G\"odel's God in
  Isabelle/HOL''.} However, the validity comes at the cost of
introducing some vagueness in the conception of Godlikeness, since the
new definition allows for there being distinct Godlike entities, which
differ only by properties that are neither positive nor non-positive.

\subsubsection{Fitting's Variant.}
Fitting suggests that there is a subtle ambiguity in G\"odel's
argument, namely, whether the notion of a \emph{positive property}
applies to extensions or intensions of properties.  In order to study
the difference, Fitting formalizes Scott's emendation in an
intensional type theory that makes it possible for him to encode and
compare both alternatives. On Fitting's interpretation, the property
of \emph{being Godlike} would be represented by the
$\lambda$-expression $[\lambda x \: \forall Y(\mathcal{P}Y \to Yx)]$,
where $\mathcal{P}$ is the second-order property of being a positive
property.  On Fitting's understanding, the variable $Y$ in the
$\lambda$-expression ranges over properties whose extensions are fixed
from world to world,\footnote{Thus, the variable $Y$ ranges over
  properties such that $\forall x(Yx \equiv \Box Yx)$.  For example
  one group of such properties can be defined in terms of an actuality
  operator: properties of the form $[\lambda x \: \mathcal{A}\varphi]$,
  i.e., \emph{being an x that is actually such that} $\varphi$, satisfy
  the condition just stated.} while $\mathcal{P}$ is a second-order
property whose extension among the first-order properties can vary
from world to world. Thus, the $\lambda$-expression that defines the
\emph{being Godlike} is a first-order property whose extension varies
from world to world.

In G\"odel's original version of the argument, positiveness and
essence are second-order properties, but Fitting suggests that the
expressions denoting the first-order properties to which positiveness
and essence apply are not rigid designators; such expressions might
have different extensions at different worlds.  So in Fitting's
variant, positiveness and essence apply only to the extensions of
first-order properties, where the expressions denoting these
extensions are rigid designators. If a property is positive at a world
$w$, its extension at every world is the same as its extension at
$w$. If we utilize the notion of a \emph{rigid property}, that is, a
property that is exemplified by exactly the same individuals in all
possible circumstances, then we can say that, on Fitting's
understanding, only rigid properties can be positive.

It should be noted that this technical notion of a positive property
departs from the ordinary notion; for example, a property like
\emph{being honest} is something a person could have in one world but
lack in another, and in those worlds where he or she has that
property, it would be considered `positive' in so far as it is
contributory to a good moral character.  But, on the above conception,
when a property like \emph{being honest} is designated a positive
property, then for any actually honest individual $x$, an alternative
world in which $x$ is not honest would be inconceivable (i.e., honesty
would be an indispensable, identity-constitutive character trait of
$x$).  In this sense, \emph{being self-identical} is a prototypical
positive property. By restricting the notions in G\"odel's argument in
this way, Fitting thus leaves Scott's variant of G\"odel's argument
largely unchanged but is able to prevent the modal collapse. This was
confirmed computationally.\footnote{Fuenmayor \& Benzm\"uller,
  ``Types, Tableaus and G\"odel's God in Isabelle/HOL''.}

\subsection{Assessment and Comparison using Ultrafilters}

These emendations proposed by Anderson and Fitting were further
investigated and assessed computationally\footnote{Benzm{\"u}ller \&
  Fuenmayor, ``Can computers help to sharpen our understanding of
  ontological arguments?''.} by extending the SSE approach in the
spirit of Fitting's book. Experiments using Isabelle/HOL that
interactively call the model finder Nitpick confirm that the formula
expressing modal collapse is not valid. The ATPs were still able to
find proofs for the main theorem not only in S5 modal logic but even
in the weaker logic KB.

In order to compare all the variant arguments by Scott, Anderson, and
Fitting, the notion of an ultrafilter was formalised in
Isabelle/HOL. On the technical level, ultrafilters were defined on the
set of rigid properties, and on the set of non-rigid, world-dependent
properties.  Moreover, in these formalizations of the variants, a
careful distinction was made between the original notion of a positive
property ($\mathcal{P}$) that applies to (intensional) properties and
a restricted notion of a positive property ($\mathcal{P'}$) that
applies to the rigidified extensions of properties that would
otherwise count as positive. Using these definitions the following
results were proved computationally:
\begin{itemize}
    \item In Scott's variant both $\mathcal{P}$ and $\mathcal{P'}$
      coincide, and both have ultrafilter properties. 
    \item In Anderson's variant $\mathcal{P}$ and $\mathcal{P'}$ do
      not coincide, and only $\mathcal{P'}$ constitutes an
      ultrafilter.
    \item In Fitting's variant, $\mathcal{P}$ is not considered an
      appropriate notion and so not defined. However, $\mathcal{P'}$
      is an ultrafilter.
\end{itemize}
Our computational experiments thus reveal an intriguing correspondence
between the variants of the ontological argument by Anderson and
Fitting, which otherwise seem quite different. The variants of
Anderson and Fitting require that only the restricted notion of a
positive property is an ultrafilter.

\subsection{Future Research}
The above insights suggest an alternative approach to the argument,
namely, one that starts out with semantically introducing
$\mathcal{P}$ or $\mathcal{P'}$ as ultrafilters and then reconstructs
variants of the formal argument on the basis of this semantics.  This
could lead to an alternative reconstruction in which some of the
axioms of the variants described above could be derived as theorems.

The experimental setup described above also provides a basis for
interesting research about how to prove that there is a \emph{unique}
object that exemplifies the property \emph{being God}. G\"odel's
original premise set guarantees that there is a unique such object,
but on pain of modal collapse. The emendations prevent the modal
collapse but at the loss of a unique object that exemplifies
\emph{being God}. So it is important to study how various notions of
equality in the context of the various logical settings described
above might help one to restore uniqueness. One particular motivation
is to assess whether different notions of equality do or don't yield
monotheism. Formal results about this issue would be of additional
interest theologically.

\section{Implementation of Object Theory}\label{AOTResults}

Whereas the last section described the analysis of \emph{philosophical
  arguments} using plain SSEs without abstraction layers, the focus of
this section is the analysis of full \emph{philosophical theories} by
using SSEs with abstraction layers. Section~\ref{example_S5} already
illustrated a simple example of this. We now examine a more
complicated case, namely, the analysis of AOT.

Though AOT has been developing and evolving since its first
publication,\footnote{Zalta, ``Abstract Objects''.} the basic idea,
namely, of distinguishing a new mode of predication and postulating a
plenitude of abstract objects using the new mode of predication, has
remained constant.  In all the publications on object theory, we find
a language containing the new mode of predication `$x$ \emph{encodes}
$F$' (`$xF$'), in which $F$ is a 1-place predicate. This new mode
extends the traditional second-order modal predicate calculus, which
is based on a single mode of predication, namely, $x_1,\ldots ,x_n$
\emph{exemplify} $F^n$ (`$F^nx_1\ldots x_n$'). The resulting language
allows complex formulas built up from the two modes of predication and
the system allows the two modes to be completely independent of one
another (neither $xF\to Fx$ nor $Fx\to xF$ is a theorem).

Using such a language (extended to include definite descriptions and
$\lambda$-expressions), the basic definitions and axioms of AOT have
also remained constant. If we start with a distinguished predicate
$E!$ to assert \emph{concreteness}, then the basic definitions and
axioms of AOT are:\begin{enumerate}
  \item[] Definition: Being ordinary is (defined as) being possibly
    concrete:\\ $O! = [\lambda x \: \Diamond E!x]$
  \item[] Axiom: Ordinary objects necessarily fail to encode
    properties.\\ $O!x \to \Box \neg \exists F(xF)$
  \item[] Definition: Being abstract is (defined as) not possibly
    being concrete.\\ $A! = [\lambda x \: \neg \Diamond E!x$]
  \item[] Axiom: If an object possibly encodes a property it
    necessarily does.\\ $\Diamond xF \to \Box xF$
  \item[] Comprehension Schema for Abstract Objects: Where $\varphi$
    is any condition on properties, there is an abstract object that
    encodes exactly the properties such that $\varphi$, i.e., $\exists
    x(A!x \:\&\: \forall F(xF \equiv \varphi))$, where $\varphi$ has
    no free occurrences of $x$.
\end{enumerate} 
In AOT, identity is not a primitive notion, and so the above
definitions and axioms are supplemented by a definition of identity
for objects and a definition of identity for properties, relations and
propositions.  The three most important definitions
are:\begin{description}
  \item[] Objects $x$ and $y$ are identical if and only if either $x$
    and $y$ are both ordinary objects that necessarily exemplify the
    same properties or $x$ and $y$ are both abstract objects that
    necessarily encode the same properties.
  \item Properties $F$ and $G$ are identical if and only if they are
    necessarily encoded by the same objects.
  \item Propositions $p$ and $q$ are identical just in case the
    properties $[\lambda x \: p]$ and $[\lambda x \: q]$ are
    identical.
\end{description}
While this basis has remained stable, other parts of the theory have
been developed and improved over the years.  The most recent round of
improvements, however, has been prompted by computational studies.
Some of these improvements have not yet been published.  Nevertheless,
we'll describe them here.

For example, in earlier versions of object theory,
$\lambda$-expressions of the form $[\lambda x_1\ldots x_n \:\varphi]$
in which $\varphi$ contained encoding subformulas were simply not
well-formed.  That's because certain well-known paradoxes of encoding
could arise for $\lambda$-expressions like $[\lambda x\:\exists F(xF
  \:\&\: \neg Fx)]$.\footnote{An instance of Comprehension for
  Abstract Objects that asserted the existence of an object that
  encodes just such a property would provably exemplify that property
  if and only if it did not.}  But Kirchner's computational
studies\footnote{Kirchner, ``{Representation and Partial Automation of
    the PLM in Isabelle/HOL}''.} showed that unless one is extremely
careful about the formation rules, such paradoxes could arise again by
constructing $\lambda$-expressions in which the matrix $\varphi$
included descriptions with embedded encoding formulas (encoding
formulas embedded in descriptions don't count as subformulas, and thus
were allowed).  A natural solution to avoid the re-emergence of
paradox is to no longer assume that all $\lambda$-expressions have a
denotation.  Given that AOT already included a free logic to handle
non-denoting descriptions, its free logic was extended to cover
$\lambda$-expressions. This allowed us to suppose that
$\lambda$-expressions are well-formed even if they include encoding
subformulas; the paradoxical ones simply don't denote.

Other changes to object theory that have come about as a result of
computational studies include:\begin{itemize}
  \item the notion of encoding was extended to $n$-ary encoding
    formulas and these allow one to define the logical notion of term
    existence directly by way of predication instead of by way of the
    notion of identity,
  \item the comprehension principle for propositions has been extended
    to cover all formulas of the language, so that even encoding
    formulas denote propositions, and
  \item the application of AOT to the theory of possible worlds, in
    which the latter are defined as abstract objects of a certain
    sort, was enhanced: the fundamental theorem for possible worlds,
    which asserts that a proposition is necessarily true if and only
    if it is true at all possible worlds, was extended to cover the
    new encoding propositions.
\end{itemize}
These will be explained further below, as we show how AOT was first
implemented computationally and how this led to refinements both of
the theory and its implementation.

\subsection{Construction of an SSE of AOT in Isabelle/HOL}

The first SSE of AOT that introduced abstraction layers can be found
in Kirchner's work.\footnote{Kirchner, ``{Representation and Partial
    Automation of the PLM in Isabelle/HOL}''.} A detailed description
of the structure of this SSE is beyond the scope of this paper,
however, we can nevertheless illustrate some of its features and the
challenges it had to overcome.

In order to construct an SSE, one has to represent the general model
structure of AOT in Isabelle/HOL.  The most general models of AOT are
\emph{Aczel models},\footnote{Zalta, ``Natural numbers and natural
  cardinals as abstract objects''.} an enhanced version of which we
now describe. Aczel models consist of a domain of \emph{Urelements}
that is partitioned into \emph{ordinary} Urelements and \emph{special}
Urelements. The ordinary Urelements represent AOT's ordinary objects,
whereas the special Urelements will act as proxies for AOT's abstract
objects and determine which properties abstract objects exemplify. In
addition, a domain of semantic possible worlds (and intensional
states) is assumed, and propositions are represented either as
intensions (i.e., functions from possible worlds to Booleans) or more
generally as hyperintensions (i.e., functions from intensional states
to intensions). This way the relations of AOT can be introduced before
specifying the full domain of AOT's individuals: AOT's relations can
be modeled as functions from Urelements to propositions (as the latter
were just represented). Since this already fixes the domain of
properties, a natural way to represent AOT's abstract objects that
validates their comprehension principle is to model them as sets of
properties (i.e., as sets of functions from Urelements to
propositions). The domain of AOT's individuals can now be represented
by the union of the set of ordinary Urelements and the set of sets of
properties. In order to define truth conditions for exemplification
formulas involving abstract objects, a mapping $\sigma$ that takes
abstract objects to special Urelements is required.\footnote{Note that
  for the model to be well-founded, the function $\sigma$ cannot be
  injective, i.e., $\sigma$ must map some distinct abstract objects to
  the same special Urelement.} With the help of this proxy function
$\sigma$, the truth conditions of AOT's atomic formulas can be defined
as follows:\begin{itemize}
  \item The truth conditions of an exemplification formula
    $F^nx_1\ldots x_n$ are determined by the proposition obtained by
    applying the function used to represent $F^n$ to the Urelements
    corresponding to $x_1,\ldots ,x_n$ (in such a way that when $x_i$
    is an abstract object, then its Urelement is the proxy
    $\sigma(x_i)$).  This yields a proposition, which can then be
    evaluated at a specific possible world (and in the
    hyperintensional case, at the designated `actual' intensional
    state).
  \item An encoding formula $xF$ is true if and only if $x$ is an
    abstract object and the function representing $F$ is contained in
    the set of functions representing $x$. An ordinary object $x$ does
    not encode any properties, so all formulas of the form $xF$ are
    false when $x$ is ordinary.\footnote{The truth conditions for
      $n$-ary encoding formulas $x_1\ldots x_nF^n$ can be defined on
      the basis of monadic encoding formulas, but this requires an
      appeal to the semantics of complex $\lambda$-expressions,
      discussed below. Hence we omit the discussion of this further
      development here. }
\end{itemize}
In earlier formulations, AOT relied heavily on the notion of a
\emph{propositional} formula, namely, a formula free of encoding
subformulas.  This notion played a role in relation comprehension:
only propositional formulas could be used to define new relations.
However, we realized that in the modal version of AOT, encoding
formulas are either necessarily true if true or necessarily false if
false. This led to the realization that in the models we had
constructed, all formulas could be assigned a proposition as
denotation; encoding formulas could denote propositions that are
necessarily equivalent to necessary truths or necessary falsehoods.
As a result, the latest (unpublished) versions of AOT have been
reformulated without the notion of a propositional formula and one of
the consequences of this move is that comprehension for propositions
can be extended to all formulas (this will be discussed further in
Section~\ref{sec:extended_proposition_comprehension}).

What remains to be defined are the denotations for AOT's complex
terms, namely, definite descriptions and
$\lambda$-predicates. Descriptions may fail to denote and, since AOT
follows Russell's analysis of definite descriptions, atomic formulas
containing a non-denoting description are treated as false. Therefore,
the embedding has to distinguish between the domain of individuals and
the domain for individual terms. The latter domain consists of the
domain of individuals plus an additional designated element that
represents non-denoting terms. If there exists a unique assignment to
$x$ for which it holds that $\varphi$, the definite description
$\imath x \varphi$ denotes this unique object. If there is no unique
such object, $\imath x \varphi$ denotes the designated element in the
domain of individual terms that represents non-denoting terms. The
truth conditions of atomic formulas can now just be lifted to the new
domain for terms, with the result that an atomic formula involving the
designated element for non-denoting terms becomes false.

In published versions of AOT, every well-formed $\lambda$-expression
was asserted to have a denotation. However, AOT now allows
$\lambda$-expressions with encoding subformulas and requires that some
of these (in particular, the paradoxical ones) don't denote. Only the
$\lambda$-expressions that denote are governed by $\beta$-Conversion.
Nevertheless, every $\lambda$-expression has to be interpreted in the
model, and the mechanism for doing this is as follows, where we
simplify by discussing only the 1-place case and where we suppose that
an ordinary object serves as its own proxy.  When the matrix $\varphi$
of the $\lambda$-expression $[\lambda x\: \varphi]$ has the same truth
conditions for all objects that have the same proxy, one can find a
function from Urelements to propositions that, when used to represent
$[\lambda x\: \varphi]$, preserves $\beta$-Conversion.\footnote{For
  example, if we make use of the $\lambda$-expressions of HOL's
  functional type theory, then we could point to $(\lambda u . \exists
  x .\: u \! =\! |x| \land \varphi'\, x)$ as such a function, where
  (a) $\varphi'$ is a function that represents the matrix $\varphi$ of
  the AOT $\lambda$-expression $[\lambda x \:\varphi]$ and maps the
  bound variable $x$ to a proposition, (b) the type of the bound
  variable $u$ is the type of Urelements, and (c) $|x|$ is the
  Urelement corresponding to $x$.} There is no such function when the
matrix has different truth conditions for objects with the same proxy,
but these are precisely the matrices for which the
$\lambda$-expressions provably fail to denote. We interpret these
$\lambda$-expressions in a manner similar to the interpretation of
non-denoting descriptions, namely, by introducing an additional domain
for relation \emph{terms} that extends the domain of relations with a
designated element for non-denoting terms. Since the condition under
which $[\lambda x\: \varphi]$ cannot denote is easy to formulate,
namely as $\exists x \exists y (\Box \forall F (Fx \equiv Fy) \land
\neg (\varphi \equiv \varphi^y_x))$, such expressions can be mapped to
this designated element.

Given the presence of non-denoting descriptions and
$\lambda$-expressions, AOT extended its free logic for descriptions to
cover all complex terms.  Note that the axioms of free logic are
usually stated in terms of a primitive notion of identity or a
primitive notion of existence ($\downarrow$) for terms, so that, for
example, the axiom for instantiating terms into universal claims can
be stated as one of the following:\begin{enumerate}
    \item[] $\forall \alpha\varphi \to (\exists \beta(\beta = \tau)
      \to \varphi^{\tau}_{\alpha})$
    \item[] $\forall \alpha\varphi \to (\tau\!\!\downarrow\: \to
      \varphi^{\tau}_{\alpha})$
\end{enumerate} 
Normally, these are equivalent formulations, since one usually defines
$\tau\!\!\downarrow \ \equiv \exists \beta(\beta = \tau)$.

However, object theory now proves this standard definition as a
theorem! As we saw above, it doesn't take identity as a primitive, but
rather defines it. Moreover, AOT does not take term existence as
primitive either, but defines it as well by cases: (a) an individual
term $\kappa$ \emph{exists} (`$\kappa\!\!\downarrow$') just in case
$\exists F\, F\kappa$, provided $F$ isn't free in $\kappa$, and (b) an
$n$-place property term $\Pi$ \emph{exists} (`$\Pi\!\!\downarrow$')
just in case $\exists x_1\ldots \exists x_n\, x_1\ldots x_n\Pi$,
provided no $x_i$ is free in $\Pi$, and (c) a 0-place proposition term
$\Pi$ \emph{exists} (`$\Pi\!\!\downarrow$') just in case $[\lambda x\:
  \Pi]\!\!\downarrow$, provided $x$ isn't free in $\Pi$. Thus, object
theory reduces existence to predication, and indeed, given its
definitions of identity, reduces identity to predication and
existence.\footnote{To see how the latter comes about (i.e., the
  reduction of identity to predication and existence), note that in a
  system like AOT, the definition of property identity stated in the
  opening paragraphs of Section 3, have to be formalized using
  metavariables and existence clauses in the definiens, so that we
  have:\begin{itemize}
    \item[] $\Pi\! =\! \Pi'\ =_\mathit{df}\ \Pi\!\!\downarrow \:\&\:
      {\Pi'}\!\!\downarrow \:\&\: \Box \forall x(x\Pi \equiv x\Pi')$
\end{itemize}   
The metavariables ensure that the definiendum $\Pi =\Pi'$ will be
provably false when either $\Pi$ or $\Pi'$ is non-denoting. Otherwise
one could argue, for non-denoting $\Pi$ and $\Pi'$, that both $x\Pi$
and $x\Pi'$ are equivalent (since both are false, given that atomic
formulas with non-denoting terms are false), and since this holds for
arbitrary $x$ and can be proved without an appeal to contingencies, it
follows that $\Box \forall x(x\Pi \equiv x\Pi')$. So without the
existence clauses, we could prove that $\Pi = \Pi'$ for any
non-denoting terms $\Pi$ and $\Pi'$.

So whereas identity claims in AOT require the existence of the terms
flanking the identity sign, this is not required in computational
implementations of other interesting logics. For example, Scott
introduces both a notion of ``identity" and ``existing identity",
where the latter corresponds to AOT's notion of identity. See
Benzm{\"u}ller \& Scott, ``Automating free logic in {HOL}''.} Given
the foregoing definitions, the claim $\tau\!\!\downarrow \ \equiv
\exists \beta(\beta = \tau)$ becomes a theorem.

\subsection{Minimal models of second-order modal AOT}

Normally, the minimal model for first-order quantified modal logic
(QML) contains one possible world and one individual, and in
second-order QML, there have to be at least two properties (one true
of everything and one false of everything).  So, a question arises:
what is the most natural axiom that forces the domain of possible
worlds to have at least two members (so as to exclude modal collapse),
and what effect does that have on the domain of properties?  We've
discovered that the axiom Zalta has proposed for this job in AOT,
namely, the assertion that $\Diamond \exists x(E!x \:\&\: \neg
\mathcal{A}E!x)$, not only forces the models to have at least 2
possible worlds, but also a minimum of 4 propositions and (given the
actuality operator $\mathcal{A}$ and the comprehension principle for
abstract objects) a minimum of 16 properties.  Proofs of these facts
are available within the system. The latter fact improved upon the
original discussion in PLM, which had asserted only that there are at
least 6 different properties.\footnote{Originally, Zalta had proved
  that $E!$ and its negation, $O!$, $A!$, $[\lambda x\: E!x \to E!x]$
  and the latter's negation, all of which exist by comprehension, were
  distinct properties.} But once properties in AOT were modeled in
Isabelle/HOL as functions from Urelements and possible worlds to
Booleans, it was recognized that there had to be at least 16 of those
functions.

The two core axioms that need to be considered for minimal models of
AOT are the modal axiom that requires the existence of a contingently
nonconcrete object already mentioned above and the comprehension axiom
for abstract objects:
\begin{itemize}
 \item[] $\Diamond \exists x (E!x\; \&\; \neg \mathcal{A} E! x)$
 \item[] $\exists x (A!x\; \&\; \forall F (xF \equiv \varphi))$
\end{itemize}
where \emph{being abstract} ($A!$) in the second axiom is defined as
\emph{not possibly being concrete}, i.e., where $A! = [\lambda x\;
  \neg \Diamond E!x]$. In particular, the first of these axioms
implies:
\begin{itemize}
 \item[] $\exists x (\Diamond E!x\; \&\; \neg \mathcal{A} E! x)$
\end{itemize}
while the second implies:\begin{itemize}
  \item[] $\exists x (\neg \Diamond E!x)$
\end{itemize}
From these consequences it follows that there are at least two
distinct individuals; let's call them $x_1$ and $x_2$.\footnote{To see
  this, instantiate these two existential claims using two individual
  variables $x_1$ and $x_2$, such that:\begin{itemize}
 \item[] $\Diamond E!x_1\; \&\; \neg \mathcal{A} E! x_1$
 \item[] $\neg \Diamond E!x_2$
\end{itemize}
To show that $x_1 \neq x_2$, we need the principle of the
substitution of identicals, which is asserted by the following
axiom:\begin{itemize}
 \item[] $\alpha = \beta \; \rightarrow \; (\varphi \rightarrow
   \varphi')$, whenever $\beta$ is substitutable for $\alpha$ in
   $\varphi$, and $\varphi'$ is the result of replacing zero or more
   free occurrences of $\alpha$ in $\varphi$ with occurrences of
   $\beta$.
\end{itemize}
Now if, for reductio, $x_1 = x_2$, then $\Diamond E!x_1 \rightarrow
\Diamond E!x_2$, but since $\neg\Diamond E!x_2$, this cannot be true,
hence $x_1 \neq x_2$. This already shows that there are at least
\emph{two individuals} in AOT.} The proof makes it clear that $x_1$ is
ordinary (i.e., $O!x_1$ = $[\lambda z \: \Diamond E!z]x_1$) and $x_2$
is abstract.  But by the construction of Aczel models, the proxy
Urelement of the abstract individual can't be the ordinary
individual. Urelements in Aczel models determine the exemplification
behaviour of individuals. So since $x_1$ \emph{exemplifies} being
ordinary while $x_2$ does not, $x_1$ and $x_2$ have to be mapped to
distinct Urelements.  Furthermore, the first statement $\exists
x(\Diamond E!x\; \&\; \neg \mathcal{A} E!x)$, implies that there are
at least two possible worlds in the Kripke semantics, namely, a
non-actual world, in which $E!x_1$ holds, and the actual world, in
which $E!x_1$ does not hold.

Recall that in our models, relations are represented as functions from
Urelements and possible worlds to Booleans.\footnote{In
  hyperintensional models, they additionally depend on an
  \emph{intensional state}, but since there is only one intensional
  state in a minimal model, this dependency can be ignored; it doesn't
  affect the size of the model.}  So, by a combinatorial argument from
the existence of two possible worlds and two Urelements, we may derive
the existence of at least $(2^2)^2 = 16$ relations in the model; each
relation has a well-defined and distinct exemplification extension.

However, this doesn't yet show, within the system, that there are at
least 16 distinct relations, but only that there are at least 16
distinct relations \emph{in our models} (we don't assume \emph{a
  priori} that our models are complete). However, we found a proof of
the existence of at least 16 relations in AOT and this is now part of
PLM.\footnote{The modal axiom $\exists x (\Diamond E!x\; \&\; \neg
  \mathcal{A} E! x)$ of AOT requires the existence of a contingently
  false proposition, namely $\exists x(E!x \:\&\:
  \neg\mathcal{A}E!x)$. Call the false proposition $q_0$ and its
  negation $\overline{q}_0$. These propositions (in the form of
  propositional properties $[\lambda x\: q_0]$ and $[\lambda x\:
    \overline{q}_0]$) were not considered when it was thought there
  were at least 6 properties. It turns out that they are provably
  distinct from the six other properties mentioned above and that
  combinations (e.g., conjunctions) of these propositional properties
  and the six properties mentioned above in fact yield 16 properties
  that are provably distinct in the system and correspond to the 16
  properties in the models.}

The foregoing discussion illustrates our research methodology: (1) we
constructed a model for the theory and conjectured that it was
complete; (2) we then analyzed the features of the model and arrived
at statements formulable within the systems AOT and its representation
in Isabelle/HOL that should be true given the model; (3) we
investigated whether these statements are indeed derivable in AOT (or
alternatively, derivable in the abstraction layer of the embedding);
and (4) we then concluded either that we had derived a new theorem
within these systems or that the model needed to be further refined.

\subsection{An Extended Theory of Propositions and Worlds}\label{sec:extended_proposition_comprehension}

One of the key challenges in constructing the first SSE of AOT was the
fact that its syntax relied heavily on the use of the notion of a
propositional formula (i.e., formulas with no encoding
subformulas). Only propositional formulas were allowed in the
construction of $n$-place complex relation terms for $n\geq
0$. $\varphi$ and $[\lambda \: \varphi]$ were designated as 0-place
relation terms only if $\varphi$ was a propositional formula. But
capturing the notion of a propositional formula in the SSE would have
increased its complexity significantly. For example, it would have
been necessary to define two versions of every connective and
quantifier, one for non-propositional formulas and one for
propositional formulas.  Instead, the SSE used one type for both kinds
of formulas, and thus one kind of connective and quantifier suffices.

However, from this it became apparent that the models used for the SSE
assigned every formula a proposition, including those formulas that
contained encoding subformulas. This suggested that AOT could be
expanded similarly. Consequently, the comprehension principle for
propositions in AOT was revised and expanded in such a way that it has
become a theorem that every formula denotes a proposition.  And once
every formula denotes a proposition, the fundamental theorem of
possible worlds becomes naturally extended to cover \emph{all}
formulas and not just propositional ones. The fundamental theorem of
possible world theory asserts that for every proposition $p$ and every
world $w$, $\Box p \equiv \forall w(w\models p)$, where $w\models p$
asserts that $p$ is true in $w$ (where this, in turn, is cashed out
as: $w$ encodes the propositional property $[\lambda x\: p]$). In
previous versions of AOT, only propositional formulas could be
substituted for $p$, since only propositional formulas denoted
propositions.  But once AOT was extended (as a result of our
computational investigations), every formula becomes substitutable for
$p$, including those with encoding subformulas.

\section{Generalizing the Cross-Fertilization}

As we see it, computer science and related disciplines like philosophy
that rely heavily on reasoning and argumentation, benefit from
interdisciplinary studies in which computational techniques are
applied. Historically, the realization that first-order theorem
provers don't capture the higher-order logic of many applied systems
created the impetus for the development of systems like Isabelle/HOL.
In this paper, we've seen that the requirements for implementing
logics and metaphysical theories has led to the development of new
methodologies for creating automated reasoning environments for
complex systems (e.g., those that are essentially higher-order,
non-classical, or have complex terms). This is especially clear in the
development of additional abstraction layers in which deductive
systems are recaptured so that only the theorems of the target system,
and no artifactual theorems, can be discovered
computationally. Abstraction layers in turn can be used as a technical
tool to analyze properties of the implementation, and in the case of
AOT, the completeness of its embedding. In our particular work, not
only did the interdisciplinary effort lead to improvements in the
computational methodologies used for modeling, but those same
methodologies led to improvements in the target metaphysical theory
being implemented.

This cross-fertilization methodology can be depicted more generally in
Figure~1. In this diagram, the cross-fertilization occurs primarily
between the various interactions that the user can have with the
front-end systems and applications.  Note that Isabelle/HOL integrates
state-of-the-art automated reasoning technology and benefits from the
constant improvements in all the systems that it integrates.
\begin{figure}[ht]
\includegraphics[width=\textwidth]{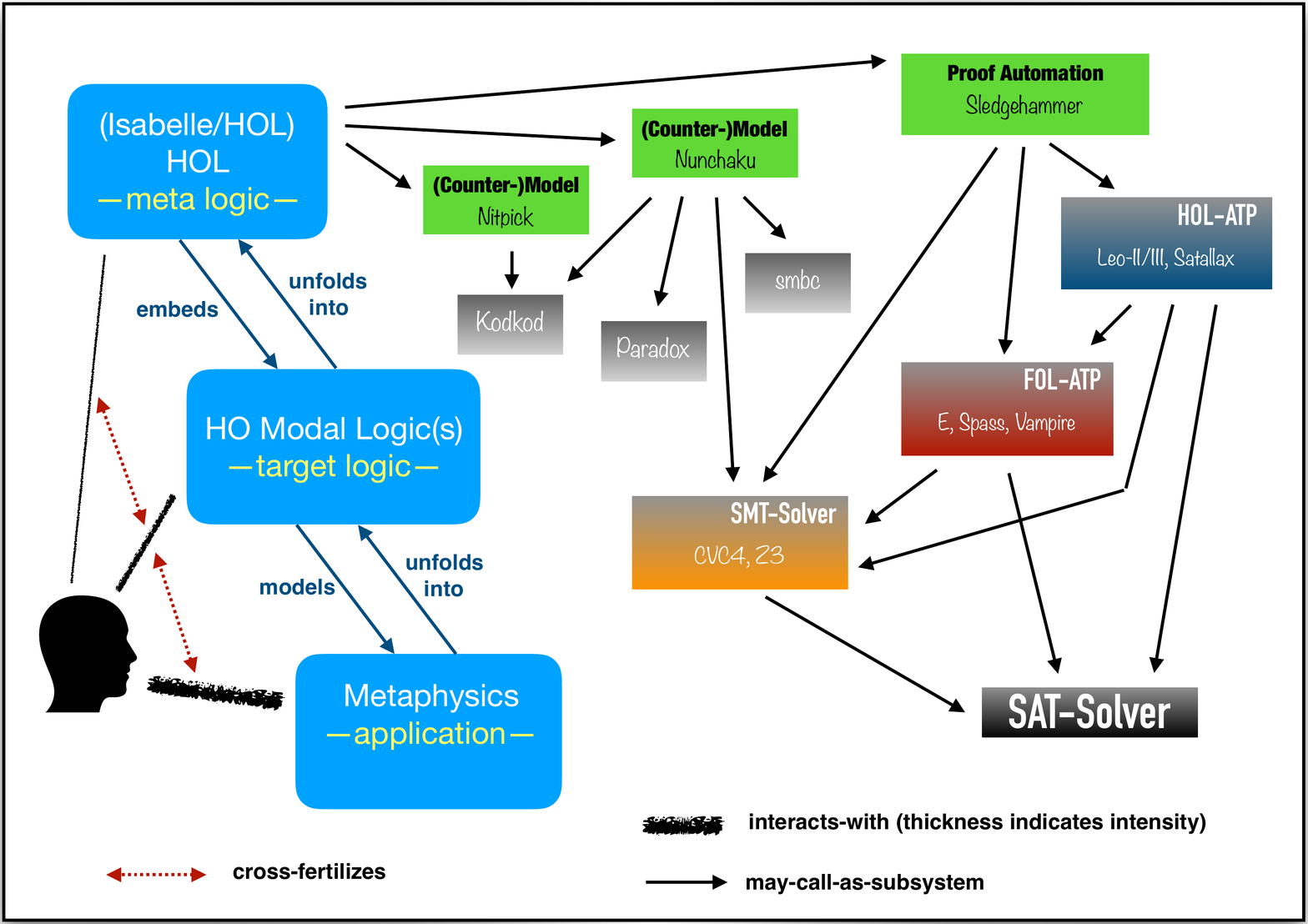} 
 \caption{Our general methodology supports the reuse of state of the
   art theorem proving technology}
\label{fig:GeneralMethodology} 
\end{figure}
In the lower left corner of Figure~1, the user is
conducting/orchestrating experiments; in this particular case, the
application in the lowest blue box is (the metaphysics of) AOT. Since
AOT is based on a higher-order modal logic, the computational
mechanization of this ``target logic" (in the middle blue box) has
served as a significant goal.  However, at the start of the project,
AOT's proof theory wasn't computationally implemented
generally. Therefore the task was to semantically embed the language
and theory in HOL (the top blue box), which turned out to be
sufficiently expressive as a meta-logic for second-order AOT. A core
advantage of this meta-logical approach is that existing reasoning
tools for HOL can readily be reused for interactive and automated
reasoning in the embedded target logic (the black arrows). This is
particularly helpful when the details of a desired language and theory
in a given context are not fully determined yet; the methodology
enables rapid prototyping of the different ways of formulating the
language and axioms of the theory.

Our preferred proof assistant for HOL has been Isabelle/HOL. This
system comes with strong user-interaction support, including a
configurable user-interface, which, in our context, enables readable
surface presentations of the embedded target logic.  Equally important
is the automation provided by the proof assistants, which include both
external ATPs orchestrated by the Sledgehammer
tool\footnote{Blanchette, B\"ohme \& Paulson ``Extending
  {Sledgehammer} with {SMT} solvers''.} and automated (counter-)model
finding tools like Nitpick\footnote{Blanchette \& Nipkow,
  ``Nitpick''.} and Nunchaku.\footnote{Cruanes \& Blanchette,
  ``Extending Nunchaku to dependent type theory''.} These systems, in
turn, make calls to specialist tools such as Kodkod, Paradox, smbc,
and the SMT solvers CVC4\footnote{Deters, Reynolds, King, Barrett \&
  Tinelli, ``A tour of CVC4''.} and Z3.\footnote{Moura \& Bj{\o}rner,
  ``Z3: An efficient SMT solver''.} Other systems integrated with
Sledgehammer include the first-order ATPs E,\footnote{Schulz, ``System
  description: {E}''.}  Spass,\footnote{Blanchette, Popescu, Wand \&
  Weidenbach, ``More {SPASS} with Isabelle''.}
Vampire,\footnote{Kov{\'{a}}cs \& Voronkov, ``First-order theorem
  proving and Vampire''.} and the higher-order ATPs
Leo-II,\footnote{Benzm{\"u}ller, Sultana, Paulson \& Thei{\ss}, ``The
  higher-order prover {LEO-II}''.} Leo-III,\footnote{Steen \&
  Benzm{\"u}ller, ``The higher-order prover {Leo-III}''.} and
Satallax.\footnote{Brown, ``Satallax''.} If one downloads
Isabelle/HOL, all of these systems are bundled with it, except for the
higher-order provers like Leo-II, Leo-III and Satallax, which can be
accessed via the TPTP infrastructure using remote calls. These
higher-order ATPs internally collaborate in turn with first-order ATPs
and SMT solvers. And all these ATPs and SMT solvers internally rely on
or integrate state-of-the-art SAT technology. Thus, whenever one of
the subsystems improves, the enhancements filter up to the
Isabelle/HOL environment.  In other words, a proof conjecture in some
theory that is not automatically solvable at the present time may well
become solvable as improvements to this framework accumulate.

When moving to other application domains (e.g., machine ethics),
deontic logics become relevant as target logics. The overall pictures
stays the same. Only the two lower blue boxes on the left of the
Figure change.  Note that the combinations of different non-classical
logics, e.g., those required for the encoding of the Gewirth principle
of generic consistency,\footnote{Fuenmayor \& Benzm\"uller, ``Alan
  Gewirth's Proof''.} can be realized and assessed as targets in this
framework.

What has been described above is a generic approach to universal
logical and metalogical reasoning\footnote{Benzm{\"u}ller, ``Universal
  (meta-)logical reasoning''.} based on shallow semantic embeddings in
HOL. In addition, the approach also supports the direct encoding of a
target logic's proof theory of choice. The shallow semantic embedding
technique and associated reasoning framework described in the previous
sections scale to applications in many other areas, including, for
example, mathematical foundations, artificial intelligence and machine
ethics. In particular, metalogical investigations are feasible beyond
what was considered possible before. In a case study in mathematics,
for example, Benzm\"uller and Scott\footnote{Benzm{\"u}ller \& Scott,
  ``Automating free logic in {HOL}''. Benzm{\"u}ller \& Scott, ``Axiom
  systems for category theory in free logic''.} compared different
axiom systems for category theory proposed by
MacLane,\footnote{MacLane, ``Groups, categories and duality''.}
Scott,\footnote{Scott, ``Identity and existence in intuitionistic
  logic''.} and Freyd \&\ Scedrov.\footnote{Freyd \& Scedrov,
  ``Categories, Allegories''.} This work started with an embedding of
free logic in HOL, which was then utilized to encode and assess the
different axiom systems. As a side result of the studies, a minor flaw
in the work of Freyd and Scedrov was revealed and
corrected. Applications in artificial intelligence include the
verification of the dependency diagrams of systems in modal
logic\footnote{Benzm{\"u}ller, Claus \& Sultana, ``Systematic
  verification of the modal logic cube''.} and an elegant,
higher-order encoding of common knowledge (of a group of agents) as
part of a solution for the wise men puzzle, a famous riddle in
artificial intelligence.\footnote{Benzm{\"u}ller, ``Universal
  (meta-)logical reasoning''.}  A normative-reasoning workbench
supporting empirical studies with alternative deontic logics that are
resistant to contrary-to-duty paradoxes is currently being
developed,\footnote{Benzm{\"u}ller, Parent \& van der Torre, ``A
  deontic logic reasoning infrastructure''.} and various embeddings of
other logics in this area can be found
elsewhere.\footnote{Benzm{\"u}ller, Farjami \& Parent, ``A dyadic
  deontic logic in {HOL}''. Benzm\"uller, Farjami \& Parent,
  ``\AA qvist's dyadic deontic logic {E} in {HOL}''. Farjami, Meder,
  Parent \& Benzm\"uller, ``{I/O} logic in {HOL}''.} A recent
extension and application of this framework\footnote{Fuenmayor \&
  Benzm\"uller, ``Alan Gewirth's Proof''.} demonstrates that even
ambitious ethical theories such as Alan Gewirth's principle of generic
consistency can be formally encoded and assessed on the computer.

\section*{References}

\begin{description}

\item J.~Alama, P.~E. Oppenheimer, and E.~N. Zalta.  \newblock
  Automating Leibniz's theory of concepts.  \newblock In A.~P. Felty
  and A.~Middeldorp, editors, {\em Automated Deduction -- {CADE-25} --
    25th International Conference on Automated Deduction, Berlin,
    Germany, August 1--7, 2015, Proceedings}, {\em Lecture Notes in
    Computer Science 9195}, pp.~73--97. Springer, 2015.

\item C.~A. Anderson.
\newblock Some emendations of {G{\"o}del's} ontological proof.
\newblock {\em Faith and Philosophy}, 7(3):291--303, 1990.

\item C.~A. Anderson and M.~Gettings.
\newblock {G\"odel}'s ontological proof revisited.
\newblock In {\em {G\"odel'96: Logical Foundations of Mathematics, Computer
  Science, and Physics: Lecture Notes in Logic 6}}, pp.~167--172. {Springer},
  1996.

\item P.~B. Andrews.  \newblock On connections and higher-order logic.
  \newblock {\em Journal of Automated Reasoning}, 5(3):257--291, 1989.

\item M.~Bentert, C.~Benzm{\"u}ller, D.~Streit, and B.~Woltzenlogel~Paleo.
\newblock Analysis of an ontological proof proposed by {Leibniz}.
\newblock In C.~Tandy, editor, {\em Death and Anti-Death, Volume 14: Four
  Decades after Michael Polanyi, Three Centuries after G.W. Leibniz}. Ria
  University Press, 2016.
\newblock Preprint URL:\\ \url{http://christoph-benzmueller.de/papers/B16.pdf}.

\item C.~Benzm{\"u}ller.
\newblock Universal (meta-)logical reasoning: Recent successes.
\newblock {\em Science of Computer Programming}, 172:48--62, March 2019.
\newblock doi:
  \url{https://doi.org/10.1016/j.scico.2018.10.008}.

\item C.~Benzm{\"u}ller, M.~Claus, and N.~Sultana.
\newblock Systematic verification of the modal logic cube in {Isabelle/HOL}.
\newblock In C.~Kaliszyk and A.~Paskevich, editors, {\em PxTP 2015}, volume
  186, pp.~27--41, Berlin, Germany, 2015. EPTCS.

\item C.~Benzm\"uller, A.~Farjami, and X.~Parent.  \newblock \AA
  qvist's dyadic deontic logic {E} in {HOL}, forthcoming;  \newblock
  \emph{Journal of Applied Logics}, preprint available online at
  \\ \url{http://orbilu.uni.lu/handle/10993/37014}.

\item C.~Benzm{\"u}ller, A.~Farjami, and X.~Parent.
\newblock A dyadic deontic logic in {HOL}.
\newblock In J.~Broersen, C.~Condoravdi, S.~Nair, and G.~Pigozzi, editors, {\em
  Deontic Logic and Normative Systems --- 14th International Conference, DEON
  2018, Utrecht, The Netherlands, 3-6 July, 2018}, pp.~33--50. College
  Publications, 2018.
\newblock (John-Jules Meyer Best Paper Award).

\item C.~Benzm{\"u}ller and D.~Fuenmayor.
\newblock Can computers help to sharpen our understanding of ontological
  arguments?
\newblock In S.~Gosh, R.~Uppalari, K.~V. Rao, V.~Agarwal, and S.~Sharma,
  editors, {\em Mathematics and Reality, Proceedings of the 11th All India
  Students' Conference on Science \& Spiritual Quest (AISSQ), 6-7 October,
  2018, IIT Bhubaneswar, Bhubaneswar, India}, pp.~195--226. The Bhaktivedanta
  Institute, Kolkata, 2018.

\item C.~Benzm{\"u}ller, X.~Parent, and L.~van~der Torre.
\newblock A deontic logic reasoning infrastructure.
\newblock In F.~Manea, R.~G. Miller, and D.~Nowotka, editors, {\em 14th
  Conference on Computability in Europe, CiE 2018, Kiel, Germany, July
  30-August, 2018, Proceedings}, {\em Lecture Notes in Computer
  Science  10936}, pp.~60--69. Springer, 2018.

\item C.~Benzm{\"u}ller and L.~Paulson.
\newblock Quantified multimodal logics in simple type theory.
\newblock {\em Logica Universalis}, 7(1):7--20, 2013.

\item C.~Benzm{\"u}ller and D.~S. Scott.
\newblock Axiom systems for category theory in free logic.
\newblock {\em Archive of Formal Proofs}, May 2018. URL = \\
\url{http://isa-afp.org/entries/AxiomaticCategoryTheory.html}

\item C.~Benzm{\"u}ller and D.~S. Scott.
\newblock Automating free logic in {HOL}, with an experimental application in
  category theory.
\newblock {\em Journal of Automated Reasoning}, 2019.
\newblock doi:\\ \url{http://dx.doi.org/10.1007/s10817-018-09507-7}.

\item C.~Benzm{\"u}ller, N.~Sultana, L.~C. Paulson, and F.~Thei{\ss}.
\newblock The higher-order prover {LEO-II}.
\newblock {\em Journal of Automated Reasoning}, 55(4):389--404, 2015.

\item C.~Benzm{\"u}ller, L.~Weber, and B.~Woltzenlogel~Paleo.
  \newblock Computer-assisted analysis of the {Anderson-H\'{a}jek}
  controversy.  \newblock {\em Logica Universalis}, 11(1):139--151,
  2017.

\item C.~Benzm{\"u}ller and B.~Woltzenlogel~Paleo.
\newblock Automating {G\"{o}del's} ontological proof of {God}'s existence with
  higher-order automated theorem provers.
\newblock In T.~Schaub, G.~Friedrich, and B.~O'Sullivan, editors, {\em ECAI
  2014}, {\em Frontiers in Artificial Intelligence and
  Applications 263}, pp.~93--98. IOS Press, 2014.

\item C.~Benzm{\"u}ller and B.~Woltzenlogel~Paleo.
\newblock Higher-order modal logics: Automation and applications.
\newblock In A.~Paschke and W.~Faber, editors, {\em Reasoning Web 2015}, 
   Lecture Notes in Computer Science: 9203, pp.~32--74, Berlin, Germany,
  2015. Springer.

\item C.~Benzm{\"{u}}ller and B.~Woltzenlogel~Paleo.  \newblock
  Interacting with modal logics in the {Coq} proof assistant.
  \newblock In L.~D. Beklemishev and D.~V. Musatov, editors, {\em
    Computer Science --- Theory and Applications --- 10th
    International Computer Science Symposium in Russia, {CSR} 2015,
    Listvyanka, Russia, July 13-17, 2015, Proceedings}, {\em Lecture
    Notes in Computer Science  9139}, pp.~398--411. Springer, 2015.

\item C.~Benzm{\"u}ller and B.~Woltzenlogel~Paleo.
\newblock An object-logic explanation for the inconsistency in {G\"odel's}
  ontological theory (extended abstract, sister conferences).
\newblock In M.~Helmert and F.~Wotawa, editors, {\em KI 2016: Advances in
  Artificial Intelligence, Proceedings}, Lecture Notes in Computer Science,
  pp.~244--250, Berlin, Germany, 2016. Springer.

\item C.~Benzm{\"u}ller and B.~Woltzenlogel~Paleo.  \newblock The
  inconsistency in {G{\"o}del's} ontological argument: A success story
  for {AI} in metaphysics.  \newblock In S.~Kambhampati, editor, {\em
    IJCAI 2016}, volume 1-3, pp.~936--942. AAAI Press, 2016.

\item C.~Benzm{\"u}ller and B.~Woltzenlogel~Paleo.
\newblock The modal collapse as a collapse of the modal square of opposition.
\newblock In J.-Y. B\'{e}ziau and G.~Basti, editors, {\em The Square of
  Opposition: A Cornerstone of Thought (Collection of papers related to the
  World Congress on the Square of Opposition IV, Vatican, 2014)}, Studies in
  Universal Logic, pp.~307--313. Springer International Publishing
  Switzerland, 2016.
\newblock \url{http://www.springer.com/us/book/9783319450612}.

\item Y.~Bertot and P.~Casteran.  \newblock {\em {Interactive Theorem
    Proving and Program Development -- Coq'Art: The Calculus of
    Inductive Constructions}}.  \newblock Texts in Theoretical
  Computer Science. Springer, 2004.

\item F.~Bj{\o}rdal.  \newblock Understanding {G\"{o}del’s}
  ontological argument.  \newblock In T.~Childers, editor, {\em The
    Logica Yearbook 1998}, pp.~214--217. Filosofia, 1999.

\item J.~C. Blanchette, S.~B\"ohme, and L.~C. Paulson.
\newblock Extending {Sledgehammer} with {SMT} solvers.
\newblock {\em Journal of Automated Reasoning}, 51(1):109--128, 2013.

\item J.~C. Blanchette and T.~Nipkow.
\newblock Nitpick: {A} counterexample generator for higher-order logic based on
  a relational model finder.
\newblock In M.~Kaufmann and L.~C. Paulson, editors, {\em Interactive Theorem
  Proving, First International Conference, {ITP} 2010, Edinburgh, UK, July
  11-14, 2010. Proceedings}, {\em Lecture Notes in Computer
  Science  6172}, pp.~131--146. Springer, 2010.

\item J.~C. Blanchette, A.~Popescu, D.~Wand, and C.~Weidenbach.
\newblock {More {SPASS} with Isabelle -- Superposition with Hard Sorts and
  Configurable Simplification}.
\newblock In L.~Beringer and A.~P. Felty, editors, {\em Interactive Theorem
  Proving --- Third International Conference, {ITP} F2012, Princeton, NJ, USA,
  August 13-15, 2012. Proceedings}, {\em Lecture Notes in
  Computer Science 7406}, pp.~345--360. Springer, 2012.

\item C.~E. Brown.
\newblock Satallax: An automatic higher-order prover.
\newblock In B.~Gramlich, D.~Miller, and U.~Sattler, editors, {\em Automated
  Reasoning --- 6th International Joint Conference, {IJCAR} 2012, Manchester,
  UK, June 26-29, 2012. Proceedings},  {\em Lecture Notes in
  Computer Science 7364}, pp.~111--117. Springer, 2012.

\item S.~Cruanes and J.~C. Blanchette.
\newblock Extending Nunchaku to dependent type theory.
\newblock In {\em Proceedings First International Workshop on Hammers for Type
  Theories, HaTT@IJCAR 2016, Coimbra, Portugal, July 1, 2016.}, volume 210 of
  {\em {EPTCS}}, pp.~3--12, 2016.

\item L.~de~Moura and N.~Bj{\o}rner.
\newblock Z3: An efficient SMT solver.
\newblock In C.~R. Ramakrishnan and J.~Rehof, editors, {\em Tools and
  Algorithms for the Construction and Analysis of Systems}, pp.~337--340,
  Berlin, Heidelberg, 2008. Springer Berlin Heidelberg.

\item M.~Deters, A.~Reynolds, T.~King, C.~W. Barrett, and C.~Tinelli.
\newblock A tour of {CVC4:} {How} it works, and how to use it.
\newblock In K.~Claessen and V.~Kuncak, editors, {\em Formal Methods in
  Computer-Aided Design, {FMCAD} 2014, Lausanne, Switzerland, October 21-24,
  2014}, page~7. {IEEE}, 2014.

\item A.~Farjami, P.~Meder, X.~Parent, and C.~Benzm\"uller.
\newblock {I/O} logic in {HOL}, forthcoming.
\newblock \emph{Journal of Applied Logics}; 
  preprint available online, URL = \\ \url{http://orbilu.uni.lu/handle/10993/37013}.

\item B.~Fitelson and E.~N. Zalta.
\newblock Steps toward a computational metaphysics.
\newblock {\em Journal Philosophical Logic}, 36(2):227--247, 2007.

\item M.~Fitting.
\newblock {\em Types, Tableaus, and {G}{\"o}del's God}.
\newblock Kluwer, 2002.

\item P.~Freyd and A.~Scedrov.
\newblock {\em Categories, Allegories}.
\newblock North Holland, 1990.

\item D.~Fuenmayor and C.~Benzm{\"u}ller.
\newblock Automating emendations of the ontological argument in intensional
  higher-order modal logic.
\newblock In {\em KI 2017: Advances in Artificial Intelligence 40th Annual
  German Conference on AI, Dortmund, Germany, September 25-29, 2017,
  Proceedings},  {\em LNAI 10505}, pp.~114--127. Springer, 2017.

\item D.~Fuenmayor and C.~Benzm\"uller.
\newblock {Types, Tableaus and G{\"o}del's God in Isabelle/HOL}.
\newblock {\em Archive of Formal Proofs}, 2017, 
 \url{https://www.isa-afp.org/entries/Types_Tableaus_and_Goedels_God.html}.

\item D.~Fuenmayor and C.~Benzm\"uller.  \newblock A case study on
  computational hermeneutics: {E.~J.~Lowe's} modal ontological
  argument.  \newblock {\em Journal of Applied Logics (special issue
    on Formal Approaches to the Ontological Argument)},
  5(7):1567--1603, 2018.

\item D.~Fuenmayor and C.~Benzmüller.
\newblock {Formalisation and Evaluation of Alan Gewirth's Proof for the
  Principle of Generic Consistency in Isabelle/HOL}.
\newblock {\em Archive of Formal Proofs}, Oct.~2018.
\newblock \url{http://isa-afp.org/entries/GewirthPGCProof.html}, Formal proof
  development.

\item K.~G\"odel.
\newblock {Appendix A. Notes in Kurt G\"odel's Hand}.
\newblock In J.~Sobel, editor, {\em Logic and Theism: Arguments for and Against
  Beliefs in God}, pp.~144--145. Cambridge University Press, 1970.

\item P.~H\'ajek.
\newblock Magari and others on {G\"odel’s} ontological proof.
\newblock In A.~Ursini and P.~Agliano, editors, {\em Logic and algebra}, page
  125–135. Dekker, New York etc., 1996.

\item P.~H\'ajek.
\newblock {Der Mathematiker und die Frage der Existenz Gottes}.
\newblock In B.~Buldt~et al., editor, {\em {Kurt G{\"o}del. Wahrheit und
  Beweisbarkeit}}, pp.~325--336. öbv \& hpt Verlagsgesellschaft mbH, Wien,
  2001.
\newblock ISBN 3-209-03835-X.

\item P.~H{\'{a}}jek.
\newblock A new small emendation of {G{\"{o}}del's} ontological proof.
\newblock {\em Studia Logica}, 71(2):149--164, 2002.

\item B.~Huffman and O.~Kuncar.  \newblock Lifting and transfer: {A}
  modular design for quotients in {Isabelle/HOL}.  \newblock In
  G.~Gonthier and M.~Norrish, editors, {\em Certified Programs and
    Proofs --- Third International Conference, {CPP} 2013, Melbourne,
    VIC, Australia, December 11-13, 2013, Proceedings}, {\em Lecture
    Notes in Computer Science 8307}, pp.~131--146. Springer, 2013.

\item D.~Kirchner.
\newblock {Representation and Partial Automation of the Principia
  Logico-Metaphysica in Isabelle/HOL}.
\newblock {\em Archive of Formal Proofs}, Sept. 2017.
\newblock \url{http://isa-afp.org/entries/PLM.html}.

\item L.~Kov{\'{a}}cs and A.~Voronkov.
\newblock {First-order theorem proving and Vampire}.
\newblock In N.~Sharygina and H.~Veith, editors, {\em Computer Aided
  Verification -- 25th International Conference, {CAV} 2013, Saint Petersburg,
  Russia, July 13-19, 2013. Proceedings},  {\em Lecture Notes in
  Computer Science 8044}, pp.~1--35. Springer, 2013.

\item E.~J. Lowe.
\newblock A modal version of the ontological argument.
\newblock In J.~P. Moreland, K.~A. Sweis, and C.~V. Meister, editors, {\em
  Debating Christian Theism}, chapter~4, pp.~61--71. Oxford University Press,
  2013.

\item S.~MacLane.
\newblock Groups, categories and duality.
\newblock {\em Proceedings of the National Academy of Sciences},
  34(6):263--267, 1948.

\item T.~Nipkow, L.~C. Paulson, and M.~Wenzel.  \newblock {\em
  {Isabelle/HOL} --- A Proof Assistant for Higher-Order Logic}, {\em
  LNCS 2283}.  \newblock Springer, 2002.

\item P.~E. Oppenheimer and E.~N. Zalta.
\newblock On the logic of the ontological argument.
\newblock {\em Philosophical Perspectives}, 5:509--529, 1991.

\item P.~Oppenheimer and E.~Zalta.
\newblock A computationally-discovered simplification of the ontological
  argument.
\newblock {\em Australasian Journal of Philosophy}, 89(2):333--349, 2011.

\item P.~E. Oppenheimer and E.~N. Zalta.
\newblock {Relations versus functions at the foundations of logic:
  Type-Theoretic Considerations}.
\newblock {\em Journal of Logic and Computation}, 21(2):351--374, 2011.

\item F.~J. Pelletier and E.~Zalta.
\newblock {How to say goodbye to The Third Man}.
\newblock {\em No\^us}, 34(2):165--202, 2000.

\item S.~Schulz.
\newblock System description: {E} 1.8.
\newblock In K.~L. McMillan, A.~Middeldorp, and A.~Voronkov, editors, {\em
  Logic for Programming, Artificial Intelligence, and Reasoning -- 19th
  International Conference, LPAR-19, Stellenbosch, South Africa, December
  14-19, 2013. Proceedings},  {\em Lecture Notes in Computer
  Science 8312}, pp.~735--743. Springer, 2013.

\item D.~S. Scott.
\newblock {Appendix B: Notes in Dana Scott's Hand}.
\newblock In J.~Sobel, editor, {\em Logic and Theism: Arguments for and Against
  Beliefs in God}, pp.~145--146. Cambridge U. Press, 1972.

\item D.~S. Scott.
\newblock Identity and existence in intuitionistic logic.
\newblock In M.~Fourman, C.~Mulvey, and D.~Scott, editors, {\em Applications of
  Sheaves: Proceedings of the Research Symposium on Applications of Sheaf
  Theory to Logic, Algebra, and Analysis, Durham, July 9--21, 1977}, 
   {\em Lecture Notes in Mathematics 752}, pp.~660--696. Springer, 1979.

\item J.~Sobel.
\newblock G\"odel's ontological proof.
\newblock In J.~J. Tomson, editor, {\em {On Being and Saying. Essays for
  Richard Cartwright}}, pp.~241--261. {MIT Press}, 1987.

\item J.~Sobel.
\newblock {\em Logic and Theism: Arguments for and Against Beliefs in God}.
\newblock Cambridge University Press, 2004.

\item A.~Steen and C.~Benzm{\"u}ller.  \newblock The higher-order
  prover {Leo-III}.  \newblock In D.~Galmiche, S.~Schulz, and
  R.~Sebastiani, editors, {\em Automated Reasoning. IJCAR 2018},
   {\em Lecture Notes in Computer Science 10900},
  pp.~108--116. Springer, Cham, 2018.

\item E.~Zalta.
\newblock {\em Abstract Objects: An Introduction to Axiomatic Metaphysics}.
\newblock D. Reidel, 1983.

\item E.~Zalta.
\newblock Logical and analytic truths that are not necessary.
\newblock {\em The Journal of Philosophy}, 85(2):57--74, 1988.

\item E.~Zalta.
\newblock {Twenty-Five basic theorems in situation and world theory}.
\newblock {\em Journal of Philosophical Logic}, 22(4):385--428, 1993.

\item E.~Zalta.  \newblock {Natural numbers and natural cardinals as
  abstract objects: a partial reconstruction of {Frege's} \emph{Grundgesetze}
  in object theory}.  \newblock {\em Journal of Philosophical Logic},
  28(6):619--660, 1999.

\end{description}

\newpage

\begin{appendix}

\section*{Appendix: G\"odel's Manuscript}
\begin{figure}[H] \centering
\fbox{\includegraphics[width=\textwidth]{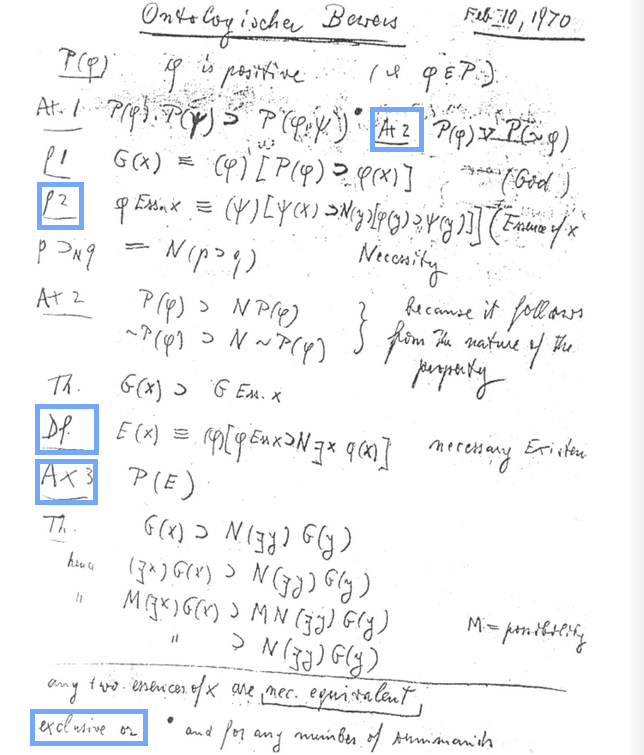}}
\caption{Page 1 of G\"odel's Manuscript.  The axioms causing the
  inconsistency in G\"odel's modal logic variant of the ontological
  argument for the existence of God are highlighted (by us) in
  blue. \textit{(Unpublished works of Kurt Gödel are Copyright
    Institute for Advanced Study and are used with permission. All
    rights reserved by Institute for Advanced
    Study.)} \label{figManuscripta}}
\end{figure}

\begin{figure}[H] \centering
\fbox{\includegraphics[width=\textwidth]{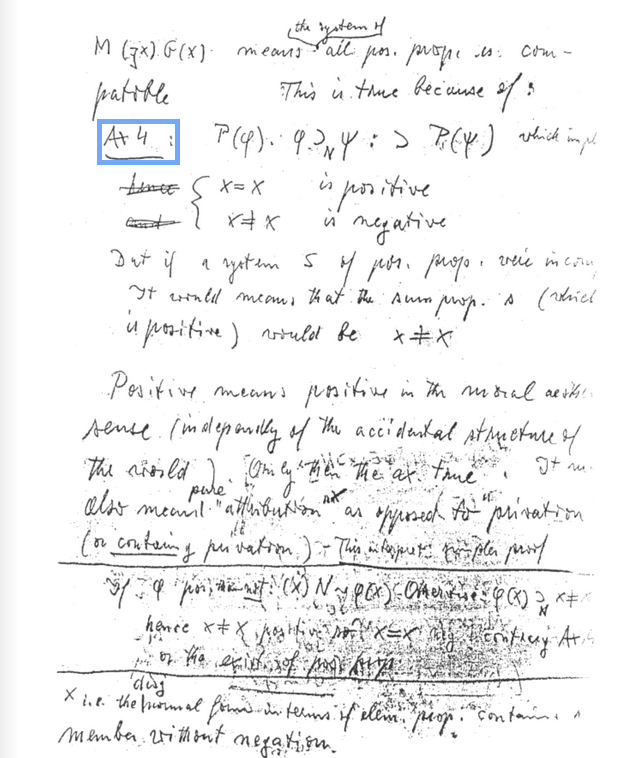}}
\caption{Page 2 of Gödel's Manuscript. The axioms causing the
  inconsistency in G\"odel's modal logic variant of the ontological
  argument for the existence of God are highlighted (by us) in
  blue. \textit{(Unpublished works of Kurt Gödel are Copyright
    Institute for Advanced Study and are used with permission. All
    rights reserved by Institute for Advanced
    Study.)} \label{figManuscriptb}}
\end{figure}

\end{appendix}

\end{document}